\documentclass[useAMS,usenatbib]{mn2e}
\usepackage{times}

\usepackage{graphicx}
\usepackage{color}
\newcommand\kms{{\rm \,km\,s^{-1}}}
\newcommand{\msun}{\rm \,M_{\odot}}
\newcommand{\kpc}{\rm \,kpc}
\newcommand{\mpc}{\rm \,Mpc}
\title[Masses for the Local Group and the Milky Way]{Masses for the Local Group
  and the Milky Way}
\author[Y.-S. Li \& S.~D.~M. White]{Yang-Shyang Li$^{1}$\thanks{Email: ysleigh@astro.rug.nl} 
and Simon D. M. White$^{2}$\thanks{Email: swhite@MPA-Garching.MPG.DE}\\
$^{1}$Kapteyn Astronomical Institute, PO Box 800, 9700 AV, Groningen, The Netherlands\\
$^{2}$Max--Planck--Institut f\"ur Astrophysik, Karl-Schwarzschild-Str. 1, D-85748 Garching, Germany}
\begin{document}

\date{Accepted. Received ; in original form }
\pagerange{\pageref{firstpage}--\pageref{lastpage}} \pubyear{2007}
\maketitle
\label{firstpage}

\begin{abstract}
We use the very large Millennium Simulation of the concordance $\Lambda$CDM
cosmogony to calibrate the bias and error distribution of Timing Argument
estimators of the masses of the Local Group and of the Milky Way. From a
large number of isolated spiral-spiral pairs similar to the Milky
Way/Andromeda system, we find the interquartile range of the ratio of timing
mass to true mass to be a factor of $1.8$, while the 5\% and 95\% points of the
distribution of this ratio are separated by a factor of $5.7$. Here we define
true mass as the sum of the ``virial'' masses $M_{200}$ of the two dominant
galaxies. For current best values of the distance and approach velocity of
Andromeda this leads to a median likelihood estimate of the true mass of the
Local Group of $5.27\times 10^{12}\msun$, or $\log M_{LG}/M_\odot = 12.72$,
with an interquartile range of $[12.58, 12.83]$ and a 5\% to 95\% range of
$[12.26, 13.01]$. Thus a 95\% lower confidence limit on the true mass of the
Local Group is $1.81\times 10^{12}\msun$. A timing estimate of the Milky Way's
mass based on the large recession velocity observed for the distant satellite
Leo I works equally well, although with larger systematic uncertainties. It
gives an estimated virial mass for the Milky Way of $2.43 \times
10^{12}\msun$ with a 95\% lower confidence limit of $0.80 \times
10^{12}\msun$.
\end{abstract}


\section{Introduction}
\label{intro_section}

During the 1970's it became generally accepted that most, perhaps all,
galaxies are surrounded by extended distributions of dark matter, so-called
dark halos \citep{eks74,opy74}. These were soon understood to play an
essential role in driving the formation and clustering of galaxies
\citep{wr78}. With the introduction of the Cold Dark Matter (CDM) paradigm,
these ideas took more concrete form, allowing quantitative predictions to be
made both for the population properties \citep{blumenthal84} and for the
large-scale clustering \citep{davis85} of galaxies. 

Measurements of the fluctuation spectrum of the Cosmic Microwave Background
\citep{smoot92,spergel03} and of the apparent acceleration of the cosmic
expansion \citep{riess98,perlmutter99} elevated the CDM model, in its variant
with a cosmological constant ($\Lambda$CDM), to the status of a standard
paradigm. At the same time improving numerical techniques and faster computers
have enabled detailed simulation of the formation and evolution of the galaxy
population within this paradigm throughout a significant fraction of the
observable Universe \citep{mr_nature}. Nevertheless, direct observational
evidence for halos as extended as the paradigm predicts around galaxies like
our own has so far come only from statistical analyses of the dynamics of
satellite galaxies \citep[e.g.][]{zaritsky97, prada03} and of the
gravitational lensing of background galaxies \citep[e.g.][]{seljak02,
mandelbaum06} based on large samples of field spirals.

The earliest observational indication that the effective mass of the Milky Way
must be much larger than its stellar mass came from the Timing Argument
(hereafter TA) of \citet{kw59}. These authors noted that the Local Group is
dominated by the two big spirals, and that these are currently approaching
each other at about $100\kms$. (The next most luminous galaxy is M33 which is
probably about a factor of $10$ less massive than M31 and the Galaxy.) This
reversal of the overall cosmic expansion must have been generated by
gravitational forces, and since the distance to the nearest external bright
galaxy is much greater than that between M31 and the Milky Way, these forces
are presumably dominated by material associated with the two spirals
themselves. 

\citeauthor{kw59} set up a simple model to analyse this situation
-- two point masses on a radial orbit.  These were at pericentre (i.e. at zero
separation) at the Big Bang and must have passed through apocentre at least
once in order to be approaching today. Clearly this requires an apocentric
separation larger than the current separation and an orbital period less than
twice the current age of the Universe. Together these requirements put a lower
limit on on the total mass of the pair. A more precise estimate of the minimum
possible mass is obtained from the parametric form of Kepler's laws for a
zero angular momentum orbit:
\begin{equation}
  r=a(1-\cos\chi)
\end{equation}

\begin{equation}
  t=\bigg(\frac{a^{3}}{GM}\bigg)^{1/2}(\chi-\sin\chi)
\end{equation}

\begin{equation}
  \frac{dr}{dt}=\sqrt{\frac{GM}{a}}\frac{\sin\chi}{1-\cos\chi}
\end{equation}
where $r$ is the current separation, $dr/dt$ is the current relative
velocity, $a$ is the semi-major axis, $\chi$ is the eccentric anomaly, $t$ is
the time since the Big Bang (the age of the universe) and $M$ is total mass
\citep{lb81}. Given observationally determined values for $r$, $dr/dt$ and
$t$, these equations have an infinite set of discrete solutions for $\chi$,
$a$ and $M$ labelled by the number of apocentric passages since the Big
Bang. The solution corresponding to a single apocentric passage gives the
smallest (and only plausible) estimate for the mass, which is about
$5\times 10^{12}\msun$ for current estimates of $r$, $dr/dt$ and $t$. Note
that this is still a lower limit on the total mass, even within the simple
point-mass binary model, since any non-radial motions in the system would
increase its present kinetic energy and so increase the mass required to
reverse the initial expansion and bring the pair to their observed separation
by the present day \citep[see][]{el82}.

As \citeauthor{kw59} realised, this timing estimate of the total mass of the Local
group exceeds by more than an order of magnitude the mass within the visible
regions of the galaxies, as estimated from their internal dynamics, in
particular, from their rotation curves. Thus 90\% of the mass must lie outside
the visible galaxies and be associated with little or no detectable light.
Modern structure formation theories like $\Lambda$CDM predict this mass to be
in extended dark matter halos with $M(r)$ increasing very roughly as $r$ out
to the point where the halos of the two galaxies meet. Such structures have no
well-defined edge, so any definition of their total mass is necessarily
somewhat arbitrary. In addition, their dynamical evolution from the Big Bang
until the present is substantially more complex than that of a point-mass
binary. Thus the mass value returned by the Timing Argument cannot be
interpreted without some calibration against consistent dynamical models with
extended dark halos. 

A first calibration of this type was carried out by \citet{kc91} using
simulations of an Einstein-de Sitter CDM cosmogony.  Here we use the very much
larger Millennium Simulation \citep{mr_nature} to obtain a more refined
calibration based on a large ensemble of galaxy pairs with observable
properties similar to those of the Local Group. We find that the standard
timing estimate is, in fact, an almost unbiased estimate of the sum of the
conventionally defined virial masses of the two large galaxies.

\citet{zaritsky89} attempted to measure the halo mass of the Milky Way alone
by measuring radial velocities for its dwarf satellites and assuming the
population to be in dynamical equilibrium in the halo potential. They noted,
however, that one of the most distant dwarfs, Leo I, has a very large
recession velocity and as a result provides a interesting lower limit on the
Milky Way's mass by a variant of the original Timing Argument. To reach its
present position and radial velocity, Leo I must have passed pericentre at
least once since the Big Bang and now be receding from the Galaxy for (at
least) the second time. 

Applying the point-mass radial orbit Equations (1) -- (3) to this case gives a
lower bound of about $1.6\times 10^{12}\msun$.  This seems likely to be a
significant underestimate, since Leo I could not have passed through the
centre of the Milky Way without being tidally destroyed so its orbit cannot be
purely radial. Below we calibrate the Timing Argument for this case also,
finding it to work well although with significantly more scatter than for the
Local Group as a whole. This is because the $\Lambda$CDM paradigm predicts
that the dynamics on the scale of Leo I's orbit ($\sim 200\kpc$) is typically
more complex than on the scale of the Local Group as a whole ($\sim 700\kpc$).

Our paper is organised as follows.  In Section~\ref{data_section}, we briefly
describe the Millennium Simulation and the selection criteria we use to define
various samples of `Local Group-like' pairs and of `Milky Way-like' halos. In
Section~\ref{result_section}, we plot the ratio of true total mass to Timing
Argument mass estimate for these samples, and we use its distribution to
define an unbiased TA estimator of true mass with its associated confidence
ranges. In Section~\ref{app_LG_section} this is then applied to the Local
Group in order to obtain an estimate its true mass with realistic
uncertainties.  Section~\ref{app_MW_section} attempts to carry out a similar
calibration for the TA-based estimate of the Galaxy's halo mass from the orbit
of Leo I. We conclude in Section~\ref{conclusions} with a summary and
brief discussion of our results.

\section{The Millennium Simulation}
\label{data_section}

The \textit{Millennium Simulation} is an extremely large cosmological
simulation carried out by the Virgo Consortium \citep{mr_nature}.  It followed
the motion of $N=2160^{3}$ dark matter particles of mass $8.6 \times 10^{8}
~h^{-1}\msun$ within a cubic box of comoving size $500~h^{-1}\mpc$.  Its
comoving spatial resolution (set by the gravitational softening) is
5~$h^{-1}\kpc$.  The simulation adopted the concordance $\Lambda$CDM model with
parameters $\Omega_{m}=0.25, \Omega_{b}=0.045, h=0.73, \Omega_{\Lambda}=0.75,
n=1$ and $\sigma_{8}=0.9$, where, as usual, we define the Hubble constant by
$H_{0}=100h\kms\mpc^{-1}$. The current age of the universe is then $13.6
\times 10^{9}$ yr. The positions and velocities of all particles were stored
at 63 epochs spaced approximately logarithmically in expansion factor at
early times and at approximately 300 Myr intervals after $z=2$. For each such
snapshot a friends-of-friends group-finder was used to locate all virialised
structures, and their self-bound substructures (subhalos) were identified
using {\small SUBFIND} \citep{swtk01}. Halos and subhalos in neighbouring
outputs were then linked in order to build formation history trees for all the
subhalos present at each time. These data are publicly available at the
Millennium release site\footnote{http://www.mpa-garching.mpg.de/millennium}.
A ``Milky Way'' halo at $z=0$ typically contains a few thousand particles and
several resolved subhalos.

Galaxy formation was simulated within these merging history trees by using
semi-analytic models to follow the evolution of the baryonic components
associated with each halo/subhalo. Processes included are radiative cooling of
diffuse gas, star formation, the growth of supermassive black holes, feedback
of energy and heavy elements from supernovae and AGN, stellar population
evolution, galaxy merging and effects due to a reionising UV background.  The
$z=0$ galaxy catalogue we analyse here corresponds to the specific model of
\citet{croton06} and details of its assumptions and parameters can be found in
that paper. Data for the galaxy population at all redshifts are available at
the Millennium web site for the updated model of \citet{deluciab07},
as well as for the independent galaxy formation model of \citet{bower06}. All
these models are tuned to fit a wide variety of data on the nearby galaxy
population, and in addition fit many (but not all!) available data at higher
redshift \citep[see, for example,][]{kw06}.  The details of the galaxy
formation modelling are not, however, important for the dynamical issues which
are the focus of our own paper.

At $z=0$ there are $18.2 \times 10^{6}$ halos/subhalos identified in the
simulation to its resolution limit of $20$ particles. The galaxy formation model
populates these with $8,394,180$ galaxies brighter than an absolute magnitude
limit of $M_{B} = -16.7$ above which the catalogue can be considered
reasonably complete.  These catalogues list a number of properties for the
halos, subhalos and galaxies which will be important for us. Galaxies are
categorised into three types according to the nature of their association with
the dark matter.  A Type 0 galaxy sits at the centre of the dominant or main
subhalo and can be considered the central galaxy of the halo itself (formally,
of the FOF group). A Type 1 galaxy sits at the centre of one of the smaller
non-dominant subhalos associated with a FOF group. Finally, a Type 2 galaxy is
associated to a specific particle and no longer has an associated subhalo
because the object within which it formed was tidally disrupted after
accretion onto a larger halo.  Such galaxies merge with the central galaxy of
their new halo after waiting for a dynamical friction time. 

Each galaxy in the catalogue has an associated ``rotation velocity''
$V_{max}$. This is the maximum of the circular velocity $V_c(r) =
(GM(r)/r)^{1/2}$ of its subhalo for Types 0 and 1; for Type 2 objects
$V_{max}$ is frozen to its value at the latest time when the galaxy still
occupied a subhalo.  Type 0 and 1 galaxies also have an associated mass
$M_{halo}$ which is the mass of the self-bound subhalo which surrounds
them. Finally, halos of Type 0 galaxies have a conventional ``virial mass''
$M_{200}$, defined as the total mass within the largest sphere surrounding
them with an enclosed mean density exceeding 200 times the critical
value. Below we will consider both $M_{halo}$ and $M_{200}$ as possible
definitions for the ``true'' masses of M31 and the Galaxy.

We use the Millennium Simulation to construct samples of mock Milky
Way/Andromeda galaxies and of mock Local Groups as follows. We begin by
identifying all Type 0 or Type 1 galaxies with characteristic ``rotation
velocity'' either in the narrow range $200 \le V_{max}< 250\kms$ or in
the wider range, $150 \le V_{max}< 300\kms$. This produces samples of
$166,090$ and $699,177$ galaxies respectively.  The exclusion of Type 2 galaxies
reduces the samples by about 5-6\% in each case, but the excluded galaxies are
in any case not plausible analogues for the Local Group giants since they are
almost all members of large groups or clusters. We also consider subsamples in
which the morphologies predicted by the semianalytic model are forced to
approximate those of M31 and the Galaxy. Specifically, we require a
bulge-to-total luminosity ratio in the range $1.2 \le M_{B,bulge}-M_{B,total}
< 2.5$ so that the disks are 2 to 9 times brighter than the bulges in the
\textit{B}-band.  This morphology cut reduces the samples in the two $V_{max}$
ranges to $62,605$ and $271,857$ galaxies respectively.

We then identify Local Group analogues in each of these four samples by
identifying isolated pairs with separations in the range of $500-1,000\kpc$
and with negative relative radial velocities. (Note that this is the true
relative velocity rather than the relative peculiar velocity, i.e. we have
added the Hubble expansion to the relative peculiar velocity and have required
the result to be negative.)  We identify isolated pairs by keeping only those
which have no ``massive'' companion, defined as a galaxy with $V_{max} \ge
150\kms$, within a sphere of 1 Mpc radius centred on the mid-point of
the binary, and no nearby cluster, defined as a halo with $M_{200}>3\times
10^{13}\msun$ within 3 Mpc of the mid-point of the binary. These cuts
ensure that the dynamics are dominated by mass associated with the two main
systems, as appears to be the case for the Local Group.  For galaxies selected
in the narrower $V_{max}$ range we then find $178$ pairs when the morphology
cut is applied and $1,128$ pairs when it is not.  For the wider $V_{max}$
range the corresponding numbers are $2,815$ pairs and $16,479$ pairs
respectively.

When calibrating the TA estimator it proves advantageous to use simulated
pairs with dynamical state quite close to that of the real Local Group.  As we
will see below, this eliminates some systems where the dominant motion is not
in the radial direction and the TA therefore significantly underestimates the
mass. We therefore make one final cut which requires the approach velocity of
the two galaxies to lie between $0.5$ and $1.5$ times the value measured for the
real system ($-130\kms$). This results in our final sets of Local
Group lookalikes. For the narrower $V_{max}$ range we end up with $117$ pairs
when the morphology cut is applied and $758$ pairs when it is not, while for
the wider $V_{max}$ range the corresponding numbers are $1,273$ pairs and
$8,449$ pairs respectively.

When we study the application of the Timing Argument to the Milky Way--Leo I
system, we consider individual galaxies from both our $V_{max}$ ranges.  We
require these to be isolated by insisting that there should be no bright/massive
companion (with luminosity exceeding 10\% of that of the host or $V_{max} >
150\kms$) closer than 700 kpc and no massive group (defined as above) closer
than 3 Mpc. This produces samples of $137,926$ and $266,229$ potential hosts in
the cases with and without the morphology cut for the wider $V_{max}$ range,
and $29,245$ and $57,816$ potential hosts for the narrower range. We then search
for Leo I analogues around these hosts by identifying companions in the
separation range 200 to 300 kpc with $V_{max}({\rm comp}) \leq 80\kms$, 
$M_{B} < -16.7$ and $V_{ra} \ge 0.7 V_{max}({\rm host})$ where
$V_{ra}$ is the relative radial velocity of the two objects and the last
condition reflects the fact that Leo I is useful for estimating the Milky
Way's mass only because its recession velocity is comparable to the Galactic
rotation velocity ($V_{ra} \sim 0.8 V_{max}({\rm host})$ for the real Leo
I--Milky Way system).  Pairs sharing the same MW-like host are excluded in the
final list.  

With these cuts we find $344$ and $896$ satellite-host pairs in the
samples with and without the morphology cut for the looser $V_{max}$ range,
and $168$ and $374$ for the tighter range.  These relatively small numbers reflect
the fact that only about 10\% of potential hosts actually have a faint
companion in this distance range which is still bright enough to be resolved,
and fewer than 5\% of these satellites are predicted to have positive
recession velocities comparable to that observed.

\section{Results}
\label{result_section}
\subsection{Calibration of the Timing Argument mass for the Local Group}

For each simulated Local Group analogue the separation and relative radial
velocity of the two galaxies can be combined with the age of the Universe
(taken to be 13.6 Gyr) to obtain a Timing Argument mass estimate $M_{TA}$
(Equations (1) to (3)). The true mass of the pair $M_{tr}$ is harder to define
because of the extended and complex mass distributions predicted by the
$\Lambda$CDM model. The mass of an individual dark halo is often taken to be
$M_{200}$ the mass within a sphere of mean density 200 times the critical
value, so a natural choice for $M_{tr}$ is the sum of $M_{200}$ for the two
galaxies. The Millennium Simulation database only lists $M_{200}$ for Type 0
galaxies, those at the centre of the main subhalo of each friends-of-friends
particle group. Many of our LG analogues lie within a single FOF group. One of
the pair is then a Type 1 galaxy, the central object of a subdominant subhalo,
and so has no listed value for $M_{200}$. In such cases we have gone back to
the particle data for the simulation in order to measure $M_{200}$ directly
also for these galaxies.  

An alternative convention is to define $M_{tr}$ as the sum of the values of
$M_{halo}$, the maximal self-bound mass of each subhalo; this is included in
the database for both Type 0 and Type 1 galaxies. In the following we use the
notation $M_{tr,200}$ and $M_{tr,halo}$ to distinguish these two
definitions. For either we can calculate the ratio of true mass to Timing
Argument estimate,
\begin{equation}
A_x = M_{tr,x}/M_{TA}, 
\end{equation}
where the suffix $x$ is $200$ or $halo$ depending on the definition adopted
for $M_{tr}$.  If the Timing Argument is a good estimator of true mass, our
samples of LG analogues should produce a narrow distribution of $A$ values.
This distribution then allows the TA mass estimate for the real Local Group to
be converted into a best estimate of its true mass, together with associated
confidence intervals.

Our preferred sets of Local Group analogues contain simulated galaxy pairs
which mimic the real system in terms of morphology, isolation, pair separation
and pair approach velocity. In addition, they require the halos of the
simulated galaxies to have $V_{max}$ values within about $\pm 10\% $ and $\pm
35\%$ of those estimated for M31 and the Galaxy for the tight and loose ranges
of $V_{max}$, respectively. In order to understand the influence of these
constraints we give results below not only for our ``best'' samples but also
for samples where the various constraints are relaxed. Thus, we consider
samples in which 1) both morphology and isolation requirements are applied
(our preferred case), 2) the isolation requirement is removed, 3) the
morphology requirement is removed, and 4) both morphology and isolation
requirements are removed. For each case, we compare results for the
two allowed ranges of $V_{max}$ and we also examine the effect
of loosening the radial velocity constraint to $V_{ra} < 0$.

\begin{figure}
  \centerline{\includegraphics[width=0.5\textwidth]{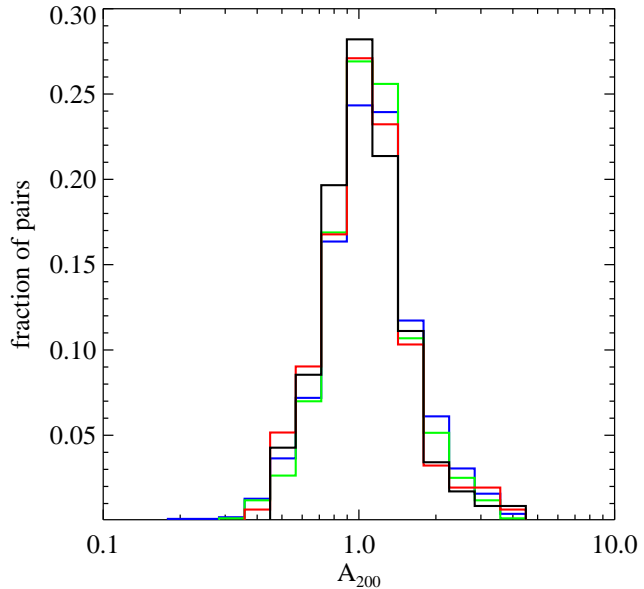}}
\caption{Normalised histograms of $A_{200}$, the ratio of true mass to
Timing Argument estimate, for samples of Local Group analogues with 
$200\kms \le V_{max} < 250\kms$ and $-195\kms<V_{ra} < -65\kms$. 
The black histogram refers to our preferred
selection where both isolation and morphology requirements are imposed. For
the red histogram the isolation requirement has been removed, for the green
histogram the morphology requirement, and for the blue histogram both
requirements.}
  \label{hist}
\end{figure}

\begin{figure}
  \centerline{\includegraphics[width=.5\textwidth]{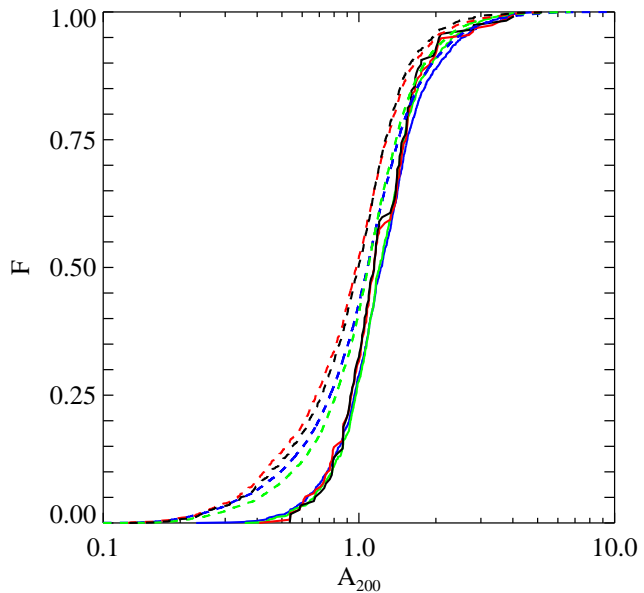}}
 \caption{Cumulative distributions of $A_{200}$ the ratio of true mass to
Timing Argument estimate for Local Group analogues with $-195\kms<
V_{ra} < -65\kms$. The solid curves correspond to the four
samples already plotted in Fig.~\ref{hist} while the dashed curves are for samples with
$150\kms \le V_{max} < 300\kms$. The colour coding is
the same as in Fig.~\ref{hist}; black indicates samples with our preferred
isolation and morphology constraints.}
  \label{narrow_ratio_cumu_dist}
\end{figure}

Fig.~\ref{hist} gives histograms of the distribution of $A_{200}$ for a sample
in the narrow $V_{max}$-range with our preferred isolation, morphology and
radial velocity cuts, as well as for three samples with the same $V_{max}$ and
$V_{ra}$ cuts but with reduced morphology and isolation requirements.
Fig.~\ref{narrow_ratio_cumu_dist} presents these same distributions in
cumulative form and compares them with the corresponding distributions for
samples with the loosened circular velocity requirement, 
$150\kms \le V_{max} < 300\kms$. In both plots black curves refer to class
(1) samples for which both isolation and morphology cuts are imposed, while
red, green and blue curves refer to samples in classes (2), (3) and (4)
respectively.  Results for the broader $V_{max}$ selection are indicated by
dashed curves in Fig.~\ref{narrow_ratio_cumu_dist}.  We give numerical results
for various percentile points of these distributions in
Table~\ref{tb_a200_vra}, and repeat all these in Table~\ref{tb_a200} for
samples where the separation velocity requirement has been loosened to $V_{ra}
< 0$.

The first and most important point to note from these these figures and tables
is that the median value of $A_{200}$ is very robust and only varies between
$0.98$ and $1.34$ for our full range of sample selection criteria. With our
preferred cuts the median values are $1.15$ and $0.99$ for the narrow and wide
$V_{max}$ samples respectively.   The best estimate of the true mass of the
Local Group (for this definition) is thus very similar to its Timing
Argument mass estimate, and depends very little on the calibrating sample of
simulated pairs.

The second important point is that the width of the distribution of $A_{200}$
does depend on how the calibrating sample is defined. In particular, it is
narrower for samples with the more restrictive $V_{max}$ range, and for given
$V_{max}$ range it is smallest for samples with our preferred cuts, those
which match the dynamical and morphological properties of the Local Group most
closely. For the narrow $V_{max}$ sample the interquartile range is a factor
of just $1.6$, and the upper and lower 5\% points are separated by a factor of
$3$. For the wider velocity range the interquartile range is a factor of $1.8$ and
the 5\% points are separated by a factor of $5.7$. This shows the Timing
Argument to be remarkably precise for systems similar to the Local Group.

The broadening of the distribution as the selection requirements are relaxed
is easy to understand. Removing the isolation requirements allows third bodies
to play a significant role in the dynamics.  This can extend the upper tail of
the $A_{200}$ distribution if mass from the third body falls inside $R_{200}$
for one of the pair galaxies or if the gravity of the third galaxy produces a
tidal field which opposes the attraction between the pair members. It can
extend the lower tail if the mass of the third body lies between the pair
members but outside their $R_{200}$ spheres, thus enhancing their mutual
attraction without adding to their mass. Removing the morphology constraint
moves the whole distribution towards larger values and this effect is most
pronounced in the large $A_{200}$ tail.  This is because objects with more
dominant bulges have more complex merger histories. They typically form in
denser regions and their halos tend to be more massive and to have more
complex structure.

Loosening the requirements on $V_{max}$ affects the distribution in a complex
way.  There is a tight relation between $V_{max}$ and $M_{200}$ (also
$M_{halo}$) in the $\Lambda$CDM structure formation model
\citep[e.g.][]{nfw97}.  Thus if we place tight restrictions on the $V_{max}$
values of our galaxies, we will obtain a sample of Local Group analogues with
a narrow range of $M_{tr}$ values. If, in addition, we force the parameters
which enter in the Timing Argument (the pair separation and relative radial
velocity) to lie in narrow ranges, then the TA mass estimate itself is tightly
constrained.  The distribution of $A_{200}$ is thus forced to be narrow as a
consequence of our selection criteria. 

A second effect is that most of the new pairs added by widening the
requirement on $V_{max}$ have at least one galaxy with $150\kms
\le V_{max} < 200\kms$, thus with relatively low $M_{tr}$. This
simply reflects the strong dependence of halo abundance on $V_{max}$. Given
that halo mass scales approximately as $V_{max}^3$ it is striking that the
addition of so many pairs containing a ``low mass'' galaxy reduces the median
value of $A_{200}$ by just 15\%.  The low tail of the distribution is more
strongly affected, by almost a factor of $2$ at the lower 5\% point, but the
upper end of the distribution is barely affected at all. This demonstrates
that the main body of the distribution is weakly affected by restrictions on
$V_{max}$, but that the lower tail (which is needed to place a lower limit on
the true mass of the Milky Way) is suppressed if $V_{max}$ is not allowed to
take small values.

\begin{table*}
 \centering
 \begin{minipage}{110mm}
  \caption{Percentage points of the $A_{200}$ distribution for samples of LG
analogues with $-195\kms< V_{ra} < -65\kms$}
  \label{tb_a200_vra}
  \begin{tabular}{@{}lcccccr@{}}
  \hline
   & 5\% & 25\% & 50\% & 75\% & 95\% & \# of pairs \\
 \hline
   $200\kms \le V_{max} < 250\kms$ & & & & & &  \\
   morphology, isolation       & 0.67 & 0.93 & 1.15 & 1.47 & 2.05  &  117 \\
   morphology, no isolation    & 0.61 & 0.93 & 1.14 & 1.52 & 2.09  &  155 \\
   no morphology, isolation    & 0.67 & 0.97 & 1.20 & 1.50 & 2.32  &  758 \\
   no morphology, no isolation & 0.63 & 0.96 & 1.22 & 1.55 & 2.54  & 1015 \\
\hline
   $150\kms \le V_{max} < 300\kms$ & & & & & &  \\
   morphology, isolation       & 0.34 & 0.72 & 0.99 & 1.27 & 1.93 & 1273 \\
   morphology, no isolation    & 0.33 & 0.68 & 0.98 & 1.29 & 2.00 & 1650 \\
   no morphology, isolation    & 0.41 & 0.81 & 1.09 & 1.40 & 2.21 & 8449 \\
   no morphology, no isolation & 0.35 & 0.77 & 1.08 & 1.43 & 2.41 &11838 \\
\hline
\end{tabular}
\end{minipage}
\end{table*}

\begin{table*}
 \centering
 \begin{minipage}{110mm}
  \caption{Percentage points of the $A_{200}$ distribution for samples of LG
analogues with $V_{ra} < 0$}
  \label{tb_a200}
  \begin{tabular}{@{}lcccccr@{}}
  \hline
   & 5\% & 25\% & 50\% & 75\% & 95\% & \# of pairs \\
 \hline
   $200\kms \le V_{max} < 250\kms$ & & & & & &  \\
   morphology, isolation       & 0.54 & 0.97 & 1.33 & 1.66 & 3.93 &  178 \\
   morphology, no isolation    & 0.45 & 0.94 & 1.26 & 1.66 & 3.93 &  241 \\
   no morphology, isolation    & 0.54 & 1.01 & 1.34 & 1.82 & 4.62 & 1128 \\
   no morphology, no isolation & 0.42 & 0.96 & 1.34 & 1.93 & 5.11 & 1596 \\
\hline
   $150\kms \le V_{max} < 300\kms$ & & & & & &  \\
   morphology, isolation       & 0.28 & 0.85 & 1.19 & 1.64 & 3.30 & 2815 \\
   morphology, no isolation    & 0.22 & 0.77 & 1.16 & 1.64 & 3.38 & 3532 \\
   no morphology, isolation    & 0.31 & 0.89 & 1.23 & 1.76 & 4.06 &16479 \\
   no morphology, no isolation & 0.18 & 0.79 & 1.19 & 1.78 & 4.46 &23429 \\
\hline
\end{tabular}
\end{minipage}
\end{table*}

In Table~\ref{tb_a200} we show the effect of weakening the cut on relative
radial velocity to require only that the two main galaxies be approaching.
Again this has remarkably little effect on the median $A_{200}$ values.  A
comparison with Table~\ref{tb_a200_vra} shows them all to be increased by
about 10\%-15\%.  The effects on the tails of the distributions are more
substantial.  The 95\% point is typically increased by about a factor of
$2$. This is because the sample now includes a substantial number of pairs with
small $V_{ra}$ (and thus smaller TA mass estimate) for which tangential motion
is important for their current orbit.  The 5\% point of the distribution is
significantly reduced, reflecting the fact that our restrictions on relative
approach velocity exclude a non negligible number of systems with approach
velocities {\it larger} than $195\kms$, and thus with large TA mass estimates.
Such systems must have more mass {\it outside} the conventional virial radii
of the two galaxies than do typical Local Group analogues in our samples.

In conclusion, we believe our most precise and robust estimate of the
distribution of $A_{200}$ to be that obtained with our preferred morphology,
isolation and radial velocity cuts for $150\kms \le V_{max} <
300\kms$, and we will use this distribution in the next section to
estimate the true mass of the Local Group.  Although the tails of the
distribution are suppressed still further for a narrower range of $V_{max}$,
this is at least in part due to the artificial effects mentioned above. In
addition the number of Local Group analogues is too small in this case for the
tails of the distribution to be reliably determined. From
Table~\ref{tb_a200_vra} we see that the best estimate of the true mass of the
Local Group (which we take to be that obtained using the median value of
$A_{200}$) is almost identical to the direct TA estimate.  The most probable
range of true mass (given by the quartiles of $A_{200}$) extends to values
about 30\% above and below this, while the 95\% confidence lower limit on the
true mass (given by the 5\% point of the $A_{200}$ distribution) is a factor
of $2.9$ smaller.

\subsection{Application to the Local Group}
\label{app_LG_section}

The three observational parameters needed to make a Timing Argument mass
estimate for the Local Group are the separation between the two main galaxies,
their radial velocity of approach and the age of the Universe. The latter is
now determined to high precision through measurements of microwave background
fluctuations. \citet{spergel07} give $13.73\pm 0.16$ Gyr. The distance to M31
is also known to high precision.  We adopt the value $784 \pm 21\kpc$ given by
\citet{sg98} based on red clump stars, noting that it agrees almost exactly
with the slightly less precise value obtained by \citet{holland98} from fits
to the colour-magnitude diagrams of M31 globular clusters. Although the
heliocentric recession velocity of M31 is known even more precisely ($-301\pm
1\kms$ according to \citealt{cvdb99}) the approach velocity of the two giant
galaxies is less certain because of the relatively poorly known rotation
velocity of the Milky Way at the Solar radius.  \citet{vdmg07} go through a
careful analysis of the uncertainties and conclude that $V_{ra} = 130\pm
8\kms$. Inserting these modern values into Equations (1) to (3) we obtain our
Timing Argument estimate of the mass of the Local Group:
\begin{equation}
   M_{LG, TA} \approx 5.32 \pm 0.48 \times 10^{12} \msun\, ,
\end{equation}
where the uncertainty is dominated by that in the relative radial
velocity. This uncertainty is still small in comparison to the scatter in the
ratio of true mass to TA estimate, so we will neglect it in the following.
The apocentric distance of the implied relative orbit of the two galaxies
is $1103\pm 30\kpc$.

We now combine this Timing Argument estimate with the distribution of
$A_{200}$ obtained in the last section for our most precise and reliable
sample of Local Group analogues (the sample with our preferred morphology,
isolation and radial velocity cuts and with the wider allowed range of
$V_{max}$) to obtain our best estimate of the true mass of the Local
Group, defined here as the sum of the $M_{200}$ values of the two
main galaxies:
\begin{equation}
   M_{LG,true} = 5.27 \times 10^{12} \msun\, ,
\end{equation}
or $\log M_{LG,true}/M_{\odot} = 12.72$. The most plausible range for this
quantity is then $[12.58, 12.83]$ with a 95\% confidence lower limit of $12.26$,
i.e $ M_{LG,true} > 1.81 \times 10^{12} \msun$ at 95\% confidence.

\subsection{Application to the Milky Way}
\label{app_MW_section}

We now calibrate the \citet{zaritsky89} Timing Argument which estimates the
mass of the Milky Way from the position and velocity of Leo I. This again
assumes a radial Keplerian orbit, but Leo I is taken to have passed through
pericentre and to be currently moving towards apogalacticon for the second
time.  Equations (1) to (3) then give a unique mass estimate $M_{MW,TA}$ for
the system for any assumed distance and radial velocity. This is taken as the
Milky Way's mass since the mass of Leo I is negligible in comparison.

Proceeding as for the Local Group, we select Milky Way -- Leo I analogues from
the Millennium Simulation in order to study the relation of this TA estimate
to the true mass of the Milky Way, which we again take to be $M_{200}$. Thus
we define the ratio
\begin{equation}
B_{200} = M_{200}/M_{MW,TA}
\end{equation}
and investigate its distribution in various samples of analogue host-satellite
systems. In particular, we consider samples of isolated host galaxies (as
defined in Section~\ref{data_section}) using both our looser and tighter
$V_{max}$ ranges, both with and without cuts on central galaxy morphology, and
requiring the distance, radial velocity and maximum circular velocity of the
satellite to satisfy $200\kpc < r < 300\kpc$, $V_{ra}\ge 0.7V_{max}({\rm host})$ and
$V_{max}({\rm comp}) \le 80\kms$.

\begin{figure}
\centerline{\includegraphics[width=.5\textwidth]{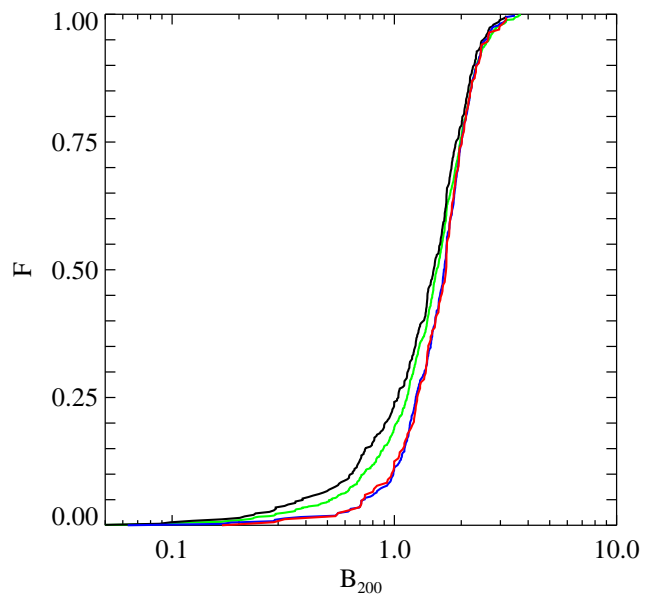}}
\caption{Cumulative distributions of $B_{200}$ the ratio of true Milky Way
mass (taken to be $M_{200}$) to TA estimate for four samples of isolated Milky
Way -- Leo I analogues from the Millennium Simulation. The red curve refers to
Milky Way analogues with $200\kms \le V_{max}({\rm host}) < 250\kms$ and
with morphology matching the Milky Way. For the black curve the circular
velocity requirement is loosened to $150\kms \le V_{max}({\rm host}) < 300\kms$, 
for the blue curve the morphology requirement is removed, and for the
green curve both requirements are relaxed. In all cases we require $V_{ra}\ge
0.7 V_{max}({\rm host})$.}
\label{B_cumul}
\end{figure}

\begin{figure*}
\includegraphics[width=.5\textwidth]{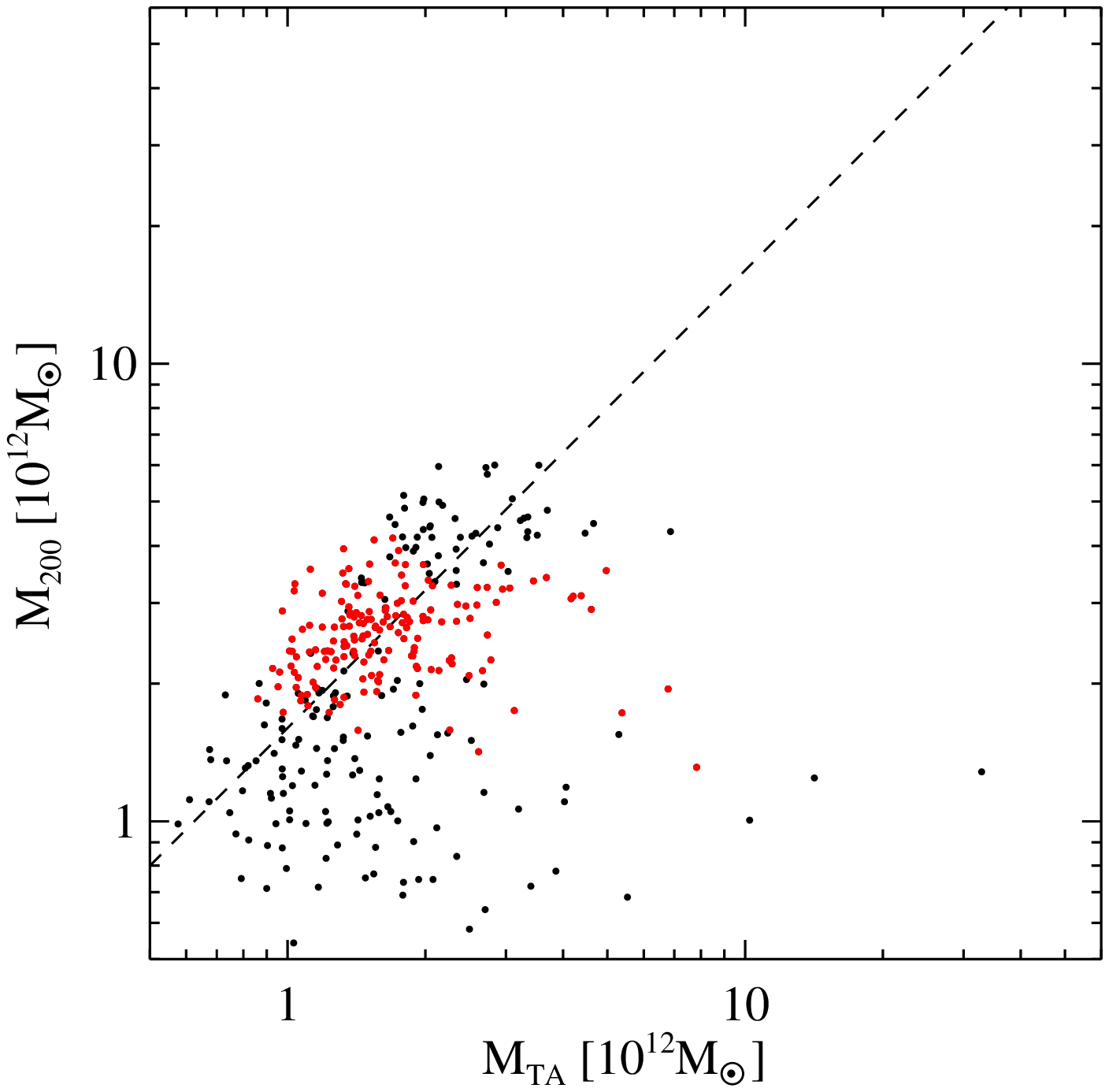}\includegraphics[width=.5\textwidth]{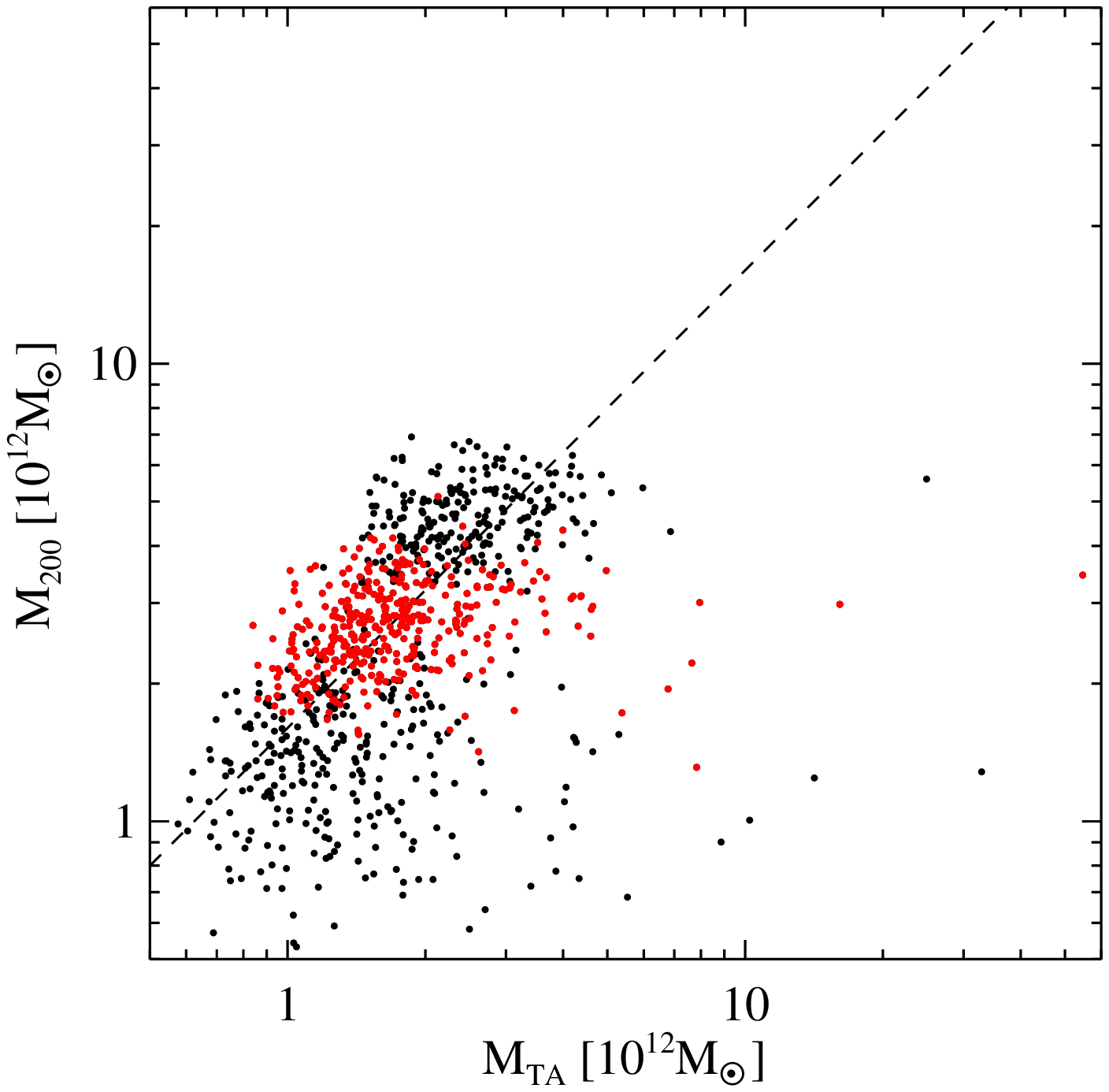}
\caption{Scatter plots of the TA estimate of the ``Milky Way's'' mass against
its true mass, $M_{200}$. The left panel is for isolated host galaxies with
morphology matching that of the real Milky Way, while the right panel is for
samples with no such morphology restriction.  In both panels red dots refer to
simulated host galaxies with $200\kms \le V_{max}({\rm host}) < 250\kms$, while black
dots indicate other hosts in the broader range $150\kms \le V_{max}({\rm host}) <
300\kms$. In all cases the analogue of Leo I is required to have $200\kpc <
r < 300\kpc$, $V_{ra}\ge 0.7 V_{max}({\rm host})$ and $V_{max}({\rm comp}) \le 80\kms$. The
straight dashed lines in the two panels show the approximate median relation
$M_{200}= 1.6 M_{MW,TA}$}
\label{mw_mass_scatter}
\end{figure*}

Results for these four samples are given in Table~\ref{tb_b200} and the
corresponding cumulative distributions of $B_{200}$ are plotted in
Fig.~\ref{B_cumul}. Scatter plots of $M_{200}$ against $M_{MW,TA}$ for the
four samples are shown in Fig.~\ref{mw_mass_scatter}.  
The behaviour is quite similar to that of the Local Group TA
mass estimator $A_{200}$. The median value of $B_{200}$ is robust and varies
very little as the definition of the analogue sample is changed. Again it is
10 -- 15\% smaller for samples with the looser $V_{max}$ selection. Unlike the
Local Group case, the median value of $B_{200}$ is about $1.6$ and so is
significantly larger than unity. This shows that $M_{MW,TA}$ is biased low as
an estimator of true Milky Way mass, reflecting the fact that tangential
motions are often significant for satellites at the distance of Leo
I. Assuming a purely radial orbit then results in an underestimate of the mass.

\begin{table*}
 \centering
 \begin{minipage}{110mm}
  \caption{Percentage points of the $B_{200}$ distribution for samples of 
  MW -- Leo I analogues with $V_{ra} \ge 0.7V_{max}({\rm host})$}
  \label{tb_b200}
  \begin{tabular}{@{}lcccccr@{}}
  \hline
   & 5\% & 25\% & 50\% & 75\% & 95\% & \# of pairs \\
 \hline
   $200\kms \le V_{max}({\rm host}) < 250\kms$ & & & & & &  \\
   morphology       & 0.71 & 1.27 & 1.71 & 2.01 & 2.62 & 168 \\
   no morphology    & 0.71 & 1.25 & 1.67 & 2.01 & 2.55 & 374 \\
\hline

   $150\kms \le V_{max}({\rm host}) < 300\kms$ & & & & & &  \\
   morphology       & 0.39 & 1.04 & 1.50 & 1.89 & 2.47 & 344 \\
   no morphology    & 0.51 & 1.14 & 1.55 & 1.98 & 2.66 & 896 \\
\hline
\end{tabular}
\end{minipage}
\end{table*}

The width of the distribution of $B_{200}$ is significantly greater for
samples with the looser $V_{max}$ selection, primarily through an extension of
the tail towards low values. This resembles the behaviour we saw above for
$A_{200}$ but it must have a different cause, since our selection criteria for
Milky Way analogues put no upper bound on $V_{ra}$, instead placing a lower
limit on $V_{ra}/V_{max}$. As a result they do not force an upper limit on
$M_{MW,TA}$ of the kind imposed on $M_{LG,TA}$ by our upper limit on $V_{ra}$
for Local Group analogues. Fig.~\ref{mw_mass_scatter} shows that the tail of
low $B_{200}$ values for the wider $V_{max}$ range is caused a relatively
small number of systems for which $M_{MW,TA}$ is anomalously large. These are
objects with anomalously large values of $V_{ra}$ and seem to occur
preferentially at small $M_{200}$, corresponding to values of $V_{max}$ below
$200\kms$.

The bulk of the points in Fig.~\ref{mw_mass_scatter} scatter fairly
symmetrically about the median relation $M_{200}= 1.6 M_{MW,TA}$ which we show
as a dashed straight line. Their mean slope is somewhat steeper than strict 
proportionality because our distance constraint on "Leo I's" is expressed in 
units of kpc rather than of $R_{200}$ or $V_{max}/H_0$.  Distant outliers 
occur only the low $M_{200}$ side
of this relation, suggesting that they may be a consequence of resolution
problems in the Millennium Simulation. For $V_{max}\sim 150\kms$, typical
halos are represented by fewer than $1,000$ particles and it seems likely that
difficulties in describing the dynamics of their satellite substructures may
begin to surface. In addition, the sample sizes are relatively small,
particularly when we impose a morphology cut, so that the estimates of the
tails of the distributions may be noisy.  This may explain in part the
apparent excess of outliers in the morphology-selected sample with the wider
$V_{max}$ range.

The observational data needed to obtain the TA estimate of the Milky Way's
mass are the age of the Universe and the Galactocentric distance and radial
velocity of Leo I. As above, we take the age of the Universe to be $13.73\pm
0.16$ Gyr from \citet{spergel07}. For the heliocentric distance to Leo I we
adopt $254\pm 19\kpc$ from \citet{bellazzini04}.  The heliocentric radial
velocity of Leo I is very precisely determined, $283\pm 0.5\kms$ according to
\citet{mom07}. Based on an assumed Galactic rotation speed at the Sun of
$220\pm 15\kms$, we derive a corresponding Galactocentric radial velocity of
$175\pm 8\kms$.  When substituted into Equations (1) to (3), these parameters
produce a TA estimate for the Milky Way's mass of
\begin{equation}
M_{MW,TA} =  1.57 \pm 0.20 \times 10^{12} \msun\, .
\end{equation}
As was the case for the Local Group, the fundamental observational quantities
are so well defined that the uncertainty of this estimate is much smaller than
the expected scatter in $B_{200}$.  We will therefore neglect it in the
following. The implied apocentric distance of Leo I is $619\pm 26\kpc$. Since
this is about half the apocentric distance of the M31 -- Milky Way relative
orbit in the TA model of Section~\ref{app_LG_section}, perturbations of the
orbit of Leo I due to the larger scale dynamics of the Local Group seem quite
likely.

For the reasons discussed above, we consider our most precise and robust
estimate for the distribution of $B_{200}$ to be that obtained for host
galaxies with $150\kms\le V_{max}({\rm host}) < 300\kms$ and with no morphology
cut. The median of this distribution then gives our best estimate of the true
halo mass of the Milky Way:
\begin{equation}
M_{200,MW} = 2.43\times 10^{12} \msun\, ,
\end{equation}
or $\log M_{200}/M_{\odot} = 12.39$. The quartiles of the distribution imply
$[12.25,12.49]$ for the most probable range of this quantity, while the 5\%
point implies a lower limit of $11.90$ at 95\% confidence. Thus the implied mass
of the Milky Way is roughly half that of the Local Group as a whole, as might
be expected on the basis of the similarity of the two giant galaxies. It is
quite similar to other recent estimates based on applying equilibrium dynamics
to the system of distant Milky Way satellites and halo stars
\citep[e.g.][]{we99,scb03}.  A significantly smaller estimate came from the
analysis of the high-velocity tail of the local stellar population by
\citet{smith07}, but we note that such analyses, in reality, only place a
lower limit on the mass of the halo, since the distribution of solar
neighbourhood stars may well be truncated at energies significantly below the
escape energy.
\subsection{An alternative mass measure?}
The halo masses we have quoted so far have been based on the ``virial masses''
$M_{200}$ of simulated halos. This choice is, of course, somewhat arbitrary,
and it may not correspond particularly well to the radii within which
individual isolated halos are approximately in static equilibrium.  As an
alternative convention, we here consider defining the mass of an individual
halo to be that of the corresponding self-bound subhalo identified by the
{\small SUBFIND} algorithm of \citet{swtk01}.  This algorithm typically
includes material outside the radius $R_{200}$ within which $M_{200}$ is
measured, but it excludes any material which is identified as part of a
smaller subhalo orbiting within the larger system. In this paper we denote
this subhalo mass as $M_{halo}$.

\begin{figure*}
  \includegraphics[width=0.5\textwidth]{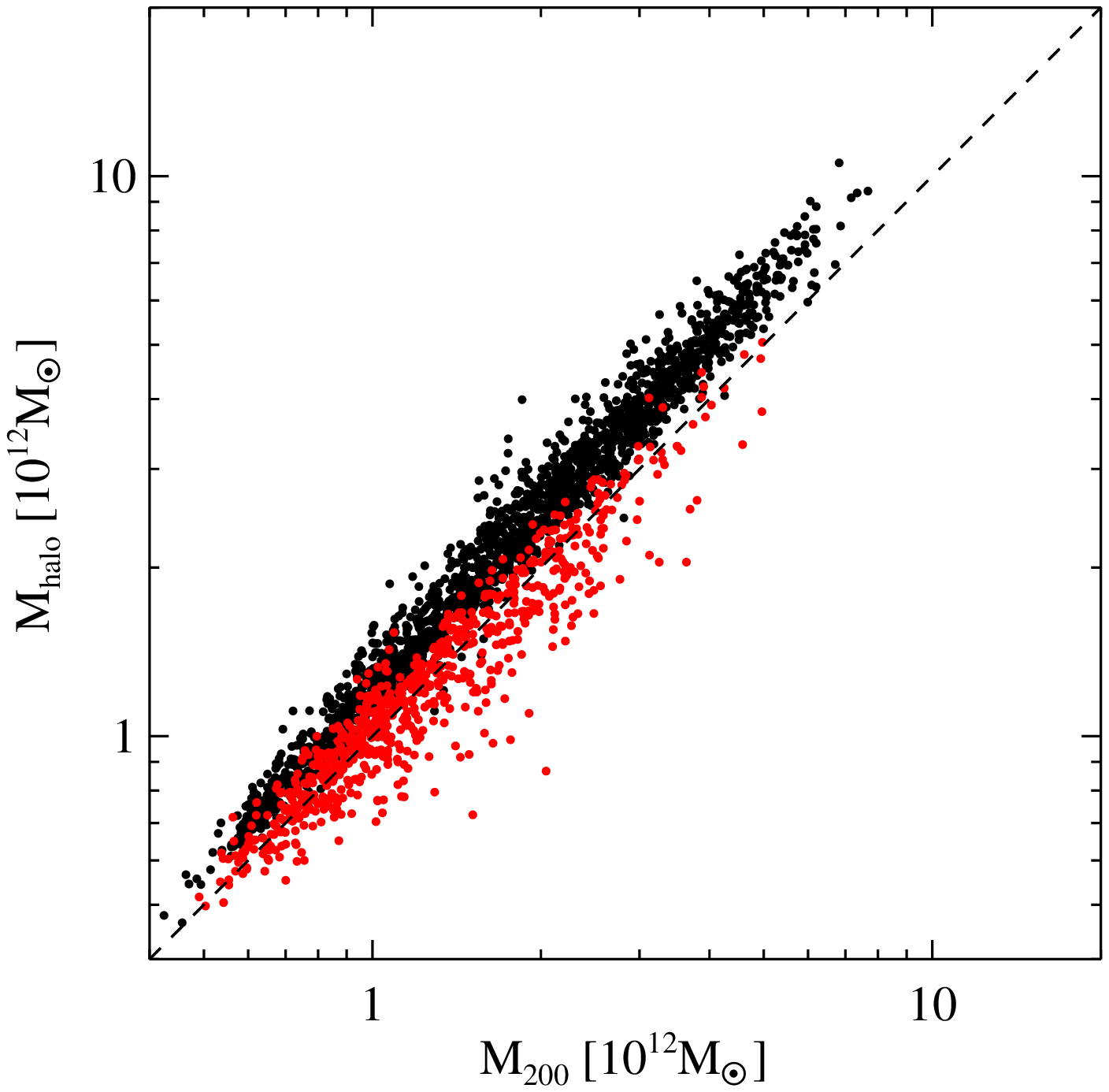}\includegraphics[width=0.5\textwidth]{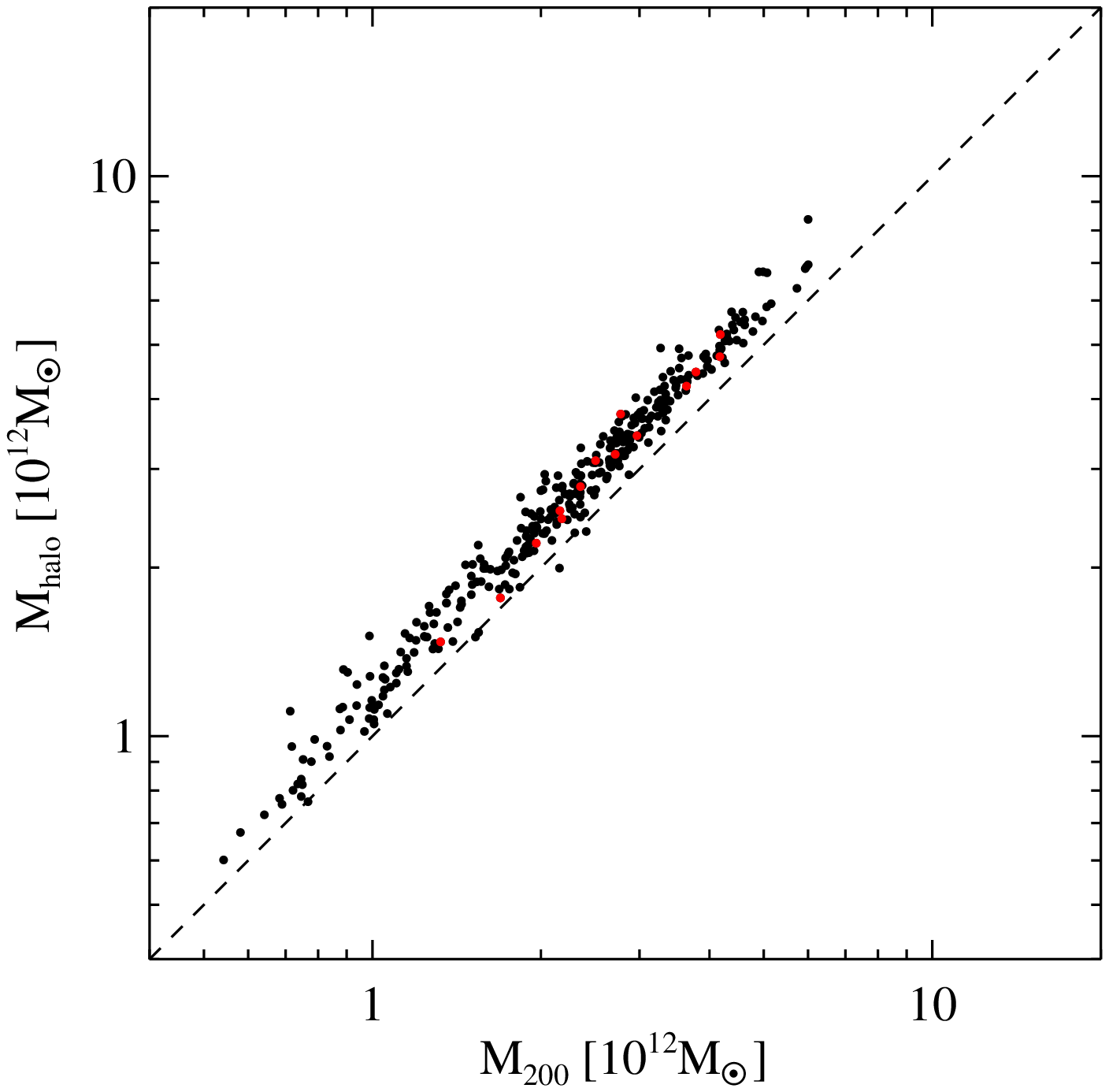}
  \caption{In the left panel we plot $M_{halo}$ against $M_{200}$ for 
all the galaxy (sub)halos in our sample of Local Group analogues with our
preferred morphology, isolation and radial velocity cuts and with $150\kms 
\le V_{max} < 300\kms$. Type 0 subhalos are plotted black
while Type 1 subhalos are red. The diagonal line is the one-to-one
relation. The right panel is a similar plot in for the ``Milky Way'' halos in
our preferred sample of Milky Way -- Leo I analogues.}
\label{mass_scatter}
\end{figure*}

In the left panel of Fig.~\ref{mass_scatter} we plot $M_{halo}$ against
$M_{200}$ for all halos in our preferred sample of Local Group analogues, that
with our preferred morphology, isolation and radial velocity cuts and with
$150\kms \le V_{max} < 300\kms$. Black and red points
in this plot refer to Type 0 and Type 1 subhalos respectively. The right panel
of Fig.~\ref{mass_scatter} is a similar plot for the ``Milky Way'' halos in
our preferred sample of Milky Way -- Leo I analogues, again the sample which
is matched in morphology and which has the wider $V_{max}$ range.  In both
panels it is clear that the correspondence between the two mass definitions is
quite tight, and that $M_{halo}$ tends to be somewhat larger than $M_{200}$.
In addition the left panel shows that Type 1 halos have smaller $M_{halo}$ for
given $M_{200}$ than do Type 0 halos, as would naively be expected.  The
average value of $\log M_{halo}/M_{200}$ for the halos in the left panel is $0.100$
for the Type 0's and $-0.004$ for the Type 1's, while it is $0.079$ for the ``Milky Way''
halos in the right panel.

This close correspondence between the two mass definitions carries over to the
distribution of our ratios of ``true'' mass to timing mass.  In
Table~\ref{tb_a200_vra_halo} we give percentage points of the $A_{halo}$
distribution for the 8 samples of Local Group analogues already considered
above.  They can be compared directly with the numbers given in
Table~\ref{tb_a200_vra} for these samples. To a good approximation the
distribution of $A_{halo}$ agrees with that of $A_{200}$ except that all
values are shifted upwards by about 16-20\%.

The same is also true for estimates of the Milky Way's mass obtained 
using the TA applied to Leo I. This can be seen from Table~\ref{tb_b200_halo} 
which repeats Table~\ref{tb_b200}
except that we now give percentage points for $B_{halo}$ rather than
$B_{200}$. Clearly it is of rather little importance which definition of halo
mass we adopt: the results obtained with our two definitions are very similar.

\begin{table*}
 \centering
 \begin{minipage}{110mm}
  \caption{Percentage points of the $A_{halo}$ distribution for samples of LG
analogues with $-195\kms< V_{ra} < -65\kms$}
  \label{tb_a200_vra_halo}
  \begin{tabular}{@{}lcccccr@{}}
    \hline
   & 5\% & 25\% & 50\% & 75\% & 95\% & \# of pairs \\
 \hline
   $200\kms \le V_{max} < 250\kms$ & & & & & &  \\
   morphology, isolation       & 0.78 & 1.11 & 1.36 & 1.72 & 2.38  &  117 \\
   morphology, no isolation    & 0.69 & 1.14 & 1.36 & 1.77 & 2.39  &  155 \\
   no morphology, isolation    & 0.76 & 1.17 & 1.44 & 1.75 & 2.74  &  758 \\
   no morphology, no isolation & 0.71 & 1.15 & 1.44 & 1.82 & 2.97  & 1015 \\
\hline
   $150\kms \le V_{max} < 300\kms$ & & & & & &  \\
   morphology, isolation       & 0.41 & 0.88 & 1.20 & 1.53 & 2.26 & 1273 \\
   morphology, no isolation    & 0.38 & 0.84 & 1.19 & 1.54 & 2.38 & 1650 \\
   no morphology, isolation    & 0.49 & 0.99 & 1.31 & 1.67 & 2.60 & 8449 \\
   no morphology, no isolation & 0.37 & 0.92 & 1.30 & 1.71 & 2.85 &11838 \\
\hline
\end{tabular}
\end{minipage}
\end{table*}

\begin{table*}
 \centering
 \begin{minipage}{110mm}
  \caption{Percentage points of the $B_{halo}$ distribution for samples of 
  MW -- Leo I analogues with $V_{ra} \ge 0.7V_{max,MW}$}
  \label{tb_b200_halo}
  \begin{tabular}{@{}lcccccr@{}}
  \hline
   & 5\% & 25\% & 50\% & 75\% & 95\% & \# of pairs \\
 \hline
   $200\kms \le V_{max} < 250\kms$ & & & & & &  \\
   morphology       & 0.90 & 1.53 & 1.98 & 2.40 & 3.14 & 168 \\
   no morphology    & 0.90 & 1.50 & 1.97 & 2.39 & 3.21 & 374 \\
\hline
   $150\kms \le V_{max} < 300\kms$ & & & & & &  \\
   morphology       & 0.42 & 1.25 & 1.81 & 2.31 & 3.09 & 344 \\
   no morphology    & 0.61 & 1.37 & 1.88 & 2.36 & 3.23 & 896 \\
\hline
\end{tabular}
\end{minipage}
\end{table*}
\section{Discussion and conclusions}
\label{conclusions}
The statistical argument underlying the analysis of this paper is more subtle
than it may at first appear, so it is worth restating it somewhat more
formally in order to understand what is being assumed in deriving the mass
estimates for the Local Group and for the Milky Way given above.

We believe that the mass distributions around galaxies are much more extended
than the visible stellar distributions, and that these have been assembled
from near-uniform ``initial'' conditions in a manner at least qualitatively
resembling that in a $\Lambda$CDM universe. Thus the assembly histories of the
Local Group and of the Milky Way's halo differ in major ways from those
assumed by the original Timing Arguments of \citet{kw59} and
\citet{zaritsky89}. In addition, the meaning of the derived mass values needs
clarification. We wish to use the Millennium Simulation to calibrate the TA
estimates against conventional measures of halo mass, and to test the general
applicability of the Timing Argument. However, we want to do this in a way
which avoids any significant dependence on the details of the $\Lambda$CDM
model, for example, on the exact density profiles, abundances and substructure
properties which it predicts for halos.

Our method uses the simulation to estimate the distribution of the ratio of
``true'' mass to TA mass estimate for samples of objects whose properties
``resemble'' those of the observed Local Group and Milky Way -- Leo I systems.
Our restrictions on separation and radial velocity implement this similarity
requirement in a straightforward way, but our constraints on $V_{max}$ have a
more complex effect. Although the true $V_{max}$ values for M31 and the Milky
Way are very likely within our looser range ($150\kms \le V_{max}
< 300\kms$) the simulation exhibits a tight correlation between
$V_{max}$ and $M_{200}$ . Imposing fixed limits on $V_{max}$ is thus
effectively equivalent to choosing a fixed range of $M_{200}$. As a result, we
are in practice estimating the distribution of $A_{200}$ or $B_{200}$ for
systems of given {\it true} mass, subject to the assumed constraints on
separation and radial velocity. However, when we apply our results to estimate
true masses for the Local Group and the Milky Way, we implicitly assume that
our distributions of $A_{200}$ and $B_{200}$ are appropriate for samples of
given TA mass estimate, again subject to our constraints on separation and
radial velocity. It is thus important to understand when these two
distributions can be considered the same.

The relation can be clarified as follows. From the simulation we compile the
distribution of $M_{tr}/M_{TA}$, or equivalently of $\Delta \equiv \ln M_{tr}
- \ln M_{TA}$, for systems with $\ln M_{tr}$ in a given range.  We then
implicitly assume that this distribution does not depend on $M_{tr}$, at least
over this range, so that the result can be taken as an estimate of the
probability density of $\Delta$ at given $M_{tr}$. Bayes Theorem then gives us
the probability density function (pdf) for $\Delta$ at fixed $M_{TA}$:
\begin{eqnarray}
f[\Delta | \ln M_{TA}] = \frac{f[\Delta, \ln M_{TA}]}{f[\ln M_{TA}]}\nonumber\\
                              = \frac{f[\Delta, \ln M_{tr}]}{f[\ln M_{TA}]} 
                                                                   \nonumber\\
                              = \frac{f[\Delta | \ln M_{tr}]\, f[\ln
                                         M_{tr}]}{ f[\ln M_{TA}]}\nonumber\\
                              = f[\Delta | \ln M_{tr}]
\label{Bayes}
\end{eqnarray}
The first line here simply writes the conditional pdf of $\Delta$ at given
$M_{TA}$ in terms of the joint pdf of the two quantities and the pdf of
$M_{TA}$. The second line then rewrites the joint pdf in terms of the
equivalent variables $\Delta$ and $M_{tr}$, using the fact that the Jacobian
of the transformation is unity. The third line re-expresses the joint pdf as
the product of the pdf of $\Delta$ at given $M_{tr}$ times the pdf of
$M_{tr}$. The final line then follows from the normalisation condition, {\it
provided} that $f[\ln M_{tr}]$ is constant and $f[\Delta | \ln M_{tr}]$ is
independent of $M_{tr}$. Thus, when estimating $M_{tr}$ from $M_{TA}$, we
assume a uniform prior on $\ln M_{tr}$ and that the distribution of $\Delta$
does not depend on true mass. Both these assumptions appear natural and
appropriate.

The analysis underlying the Timing Argument (Equations (1) -- (3)) assumes that
the relative orbit of the two objects is bound and has conserved energy since
the Big Bang. Recently, \citet{sales07} have shown that in $\Lambda$CDM models
this assumption is significantly violated for a substantial number of
satellites within halos comparable to that of the Milky Way. In particular,
they demonstrate the presence of a tail of unbound objects which are being
ejected from halos as a result of $3$-body ``slingshot'' effects during their
first pericentric passage. These objects are typically receding rapidly from
their ``Milky Way'', as assumed in the \citet{zaritsky89} argument, but they
violate its assumption that the present orbital energy can be used to infer
the period of the initial orbit (i.e. the time from the Big Bang to first
pericentric passage).  Clearly such objects should also be present in the
Millennium Simulation, although lack of resolution might make them
under-represented in comparison to the simulations analysed by
\citet{sales07}. Thus our analysis takes the possibility of such ejected
satellites into account, at least in principle. Objects of this type will show
up as systems with anomalously large TA estimates for their halo mass, and
Fig.~\ref{mw_mass_scatter} shows a number of outliers which could well be
explained in this way. Issues of this kind do not effect TA-based estimates of
the mass of the Local Group since the two big galaxies are currently
approaching for the first time.

\begin{figure}
  \includegraphics[width=0.5\textwidth]{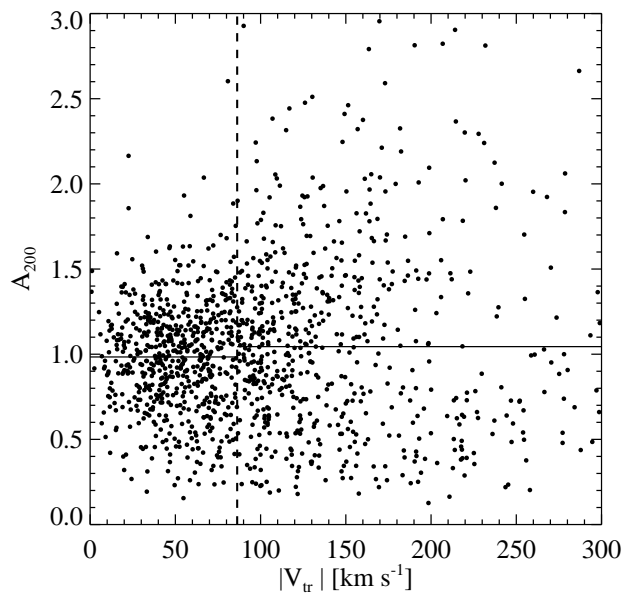} \caption{Scatter
  plot of $A_{200}$ versus transverse velocity for Local Group
  analogues in our sample with preferred morphology, isolation and
  radial velocity cuts and with $150\kms \le V_{max} < 300\kms$.  The
  vertical dashed line indicates the median $V_{tr}$.  The
  distributions on either side of this line are each further split in
  half at the median values of $V_{tr}$ (the solid horizontal lines.)
  There is essentially no correlation in this plot, indicating that a
  measurement of the transverse velocity will not significantly
  improve the TA mass estimate.}
\label{a200_vtr_scatter}
\end{figure}

The only kinematic information about the relative orbit of M31 and the
Milky Way used in our analysis is their current approach velocity.
\citet{vdmg07} show that geometric arguments can already constrain the
transverse component also, and future astrometry missions such as SIM
might allow $V_{tr}$ to be measured directly. Thus it is interesting
to ask if our TA mass estimate could be significantly refined by
measuring the full 3-D relative motion of the two galaxies, rather
than just its radial component. We address this in
Fig.~\ref{a200_vtr_scatter} which plots $A_{200}$, the ratio of true
mass to TA estimate, against $V_{tr}$ for a sample of Local Group
analogues with our preferred morphology, isolation and radial velocity
cuts, and with $150\kms \le V_{max} < 300\kms$. The median $V_{tr}$
for this sample is $86 \kms$, comparable to the \citet{vdmg07}
estimate for the real system. There is no apparent correlation of
$A_{200}$ with $V_{tr}$, and indeed, the medians of $A_{200}$ for the
high and low $V_{tr}$ halves of the sample are both close to 1 and do
not differ significantly.  Pairs with high $V_{tr}$ do show larger
{\it scatter} in $A_{200}$ than pairs on near-radial orbits. For given
separation, radial velocity and age, the Kepler model implies a mass
which increases monotonically with $V_{tr}$. The absence of a
detectable trend in Fig.~\ref{a200_vtr_scatter} shows that
uncertainties in $V_{tr}$ do not dominate the scatter in our TA mass
estimate, and that a measurement of $V_{tr}$ will not substantially
increase the precision with which the true mass can be measured.

In conclusion, our analysis shows the Timing Argument to produce robust
estimates of true mass both for the Local Group and for the Milky Way, as long
as ``true mass'' is understood to mean the sum of the conventional masses of
the major halos. For the Local Group as a whole, the estimate and confidence
limits given in Section~\ref{app_LG_section} and in the Abstract appear
reliable given the excellent statistics provided by the Millennium Simulation,
the lack of any substantial dependence on our isolation and morphology cuts,
and the relatively simple dynamical situation. Although the results based on
Leo I's orbit also appear statistically sound, the more complex dynamical
situation offers greater scope for uncertainty, particularly when trying to
place a lower limit on the mass of the Milky Way's halo.  On the other hand,
our best estimate of this mass is just under half of our estimate of the sum
of the halo masses of M31 and the Galaxy. Thus the picture presented by the
data appears quite consistent, and gives no reason to be suspicious of the
Milky Way results.
\section*{Acknowledgements}
We thank Andreas Faltenbacher for help in calculating the $M_{200}$ values of
Type 1 halos; the referee, John Dubinski, for useful suggestions regarding the transverse velocity.  YSL also thanks Martin C. Smith, Gabriella De Lucia and Amina
Helmi for useful discussions, and the Max Planck Institute for Astrophysics
for support during her visits there.  SW thanks the Kapteyn Astronomical
Institute for support as Blaauw Lecturer during extended visits to Groningen.

\label{lastpage}

\end{document}